# Thermal noise informatics: Totally secure communication via a wire; Zero-power communication; and Thermal noise driven computing

Laszlo B. Kish[(+)], Robert Mingesz[(x)], Zoltan Gingl[(x)]

[(+)]Texas A&M University, Department of Electrical and Computer Engineering, College Station, TX 77843-3128, USA

[(x)]University of Szeged, Department of Experimental Physics, Dom ter 9, Szeged, H-6720, Hungary.

## ABSTRACT

Very recently, it has been shown that thermal noise and its artificial versions (Johnson-like noises) can be utilized as an information carrier with peculiar properties therefore it may be proper to call this topic *Thermal Noise Informatics*. *Zero Power (Stealth) Communication*, *Thermal Noise Driven Computing*, and *Totally Secure Classical Communication* are relevant examples. In this paper, while we will briefly describe the first and the second subjects, we shall focus on the third subject, the secure classical communication via wire. This way of secure telecommunication utilizes the properties of Johnson(-like) noise and those of a simple Kirchhoff's loop. The communicator is unconditionally secure at the conceptual (circuit theoretical) level and this property is (so far) unique in communication systems based on classical physics. The communicator is superior to quantum alternatives in all known aspects, except the need of using a wire. In the idealized system, the eavesdropper can extract zero bit of information without getting uncovered. The scheme is naturally protected against the *man-in-the-middle attack*. The communication can take place also via currently used power lines or phone (wire) lines and it is not only a point-to-point communication like quantum channels but network-ready. We report that a pair of Kirchhoff-Loop-Johnson(-like)-Noise communicators, which is able to work over variable ranges, was designed and built. Tests have been carried out on a model-line with ranges beyond the ranges of any known direct quantum communication channel and they indicate unrivalled signal fidelity and security performance. This simple device has single-wire secure key generation/sharing rates of 0.1, 1, 10, and 100 bit/second for copper wires with diameters/ranges of 21 mm / 2000 km, 7 mm / 200 km, 2.3 mm / 20 km, and 0.7 mm / 2 km, respectively and it performs with 0.02% raw-bit error rate (99.98 % fidelity). The raw-bit security of this practical system significantly outperforms raw-bit quantum security. Current injection breaking tests show zero bit eavesdropping ability without setting on the alarm signal, therefore no multiple measurements are needed to build an error statistics to detect the eavesdropping as in quantum communication. Wire resistance based breaking tests of Bergou-Scheuer-Yariv type give an upper limit of eavesdropped raw bit ratio is 0.19 % and this limit is inversely proportional to the sixth power of cable diameter. Hao's breaking method yields zero (below measurement resolution) eavesdropping information.

## 1. INTRODUCTION

Very recently, it has been shown that thermal noise and its artificial versions (Johnson-like noises) can be utilized as an information carrier with peculiar properties therefore it may be proper to call this topic *Thermal Noise Informatics*. *Thermal Noise Driven Computing*, *Zero Power Communication*, and *Totally Secure Classical Communication* are relevant examples. In this paper, while we will briefly describe the first and the second subjects, we shall focus on the third subject, the secure communication via wire.

## 2. Zero Power Classical Communication, Zero-Quantum Quantum communication, Stealth Communication

Recently, it has been shown that the equilibrium thermal noise in information channels can be utilized to carry information. In this case, the transmitter does not emit any signal energy into the channel however it only modulates the existing noise there. This issue is completely different from the earlier Porod-Landauer debate about the question if communication without net energy cost is possible by gaining back the energy spent in the communicator devices. (Porod is right, energy-free communication is impossible just like energy-free computing). In our system, the noise is used as information carrier and no effort is made to restore the energy dissipated in the communicator devices. Therefore, this communicator is *not energy-free communication* but it is free of emitted signal energy.

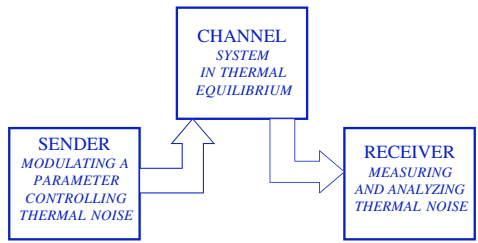

Figure 1.

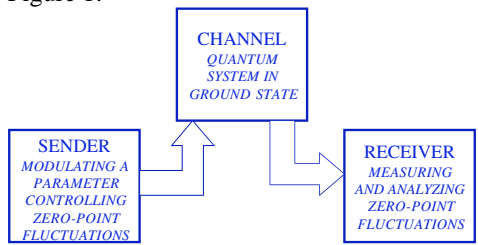

Figure 1.

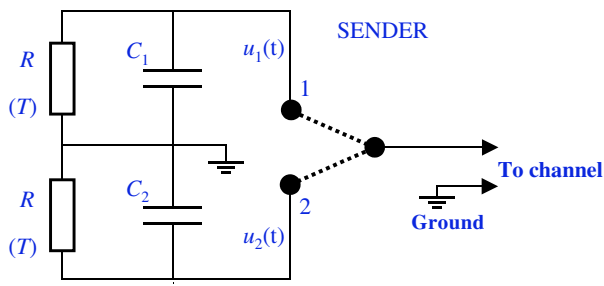

Figure 3. A possible realization of stealth communication with zero power classical communication or zero-quantum quantum communication utilizing classical thermal noise and zero-point quantum fluctuations with bandwith modulation. Classical or quantum, it depends on the upper cutoff frequency $f_c$ and the temperature. Classical: $hf_c \ll kT$; quantum: $hf_c \ll kT$.

## 3. THERMAL NOISE DRIVEN COMPUTING?

Because thermal noise turned out to be a special information carrier with low energy density in the information channel, it is natural to pose the question: can we use it for information processing and computing? At thermal noise driven computing, an idea further inspired by the fact that the neural signals in the brain are noise, certain statistical parameters of the thermal noise is the information carrier.

The information may be carried by the bandwidth as at the zero power communication example above or in other way. To study the minimal energy requirement of requirement/dissipation, another specific realization was used, a simple digital system with zero threshold voltage, when the thermal noise is equal to or greater than the digital signal. Under these conditions, when the digital signal amplitude is less than the variance of the noise, classical digital information is usually considered to be zero or useless.

We would like to note that in a different class of efforts many notable scientists, including John von Neumann [11]; Forshaw and coworkers [12,13]; and Palem and coworkers [14,15] have been proposing efficient ways of working with probabilistic switches, which are noisy digital logic units with relatively high error probability. For example, Palem and coworkers have pointed out that this may be a way to reduce power dissipation [14,15] if we give up our requirement of data accuracy. However, though these approaches can be relevant to future developments of thermal noise driven computing, they are very different from our present approach. In a *thermal noise driven computer*, the voltage in the channel is dominantly noise and the modulation of the statistical properties of this noise carry the information. Therefore, the way of extracting the information is not by "*error correcting the information*" like the efforts mentioned above do but by "*decoding the information in the noise*". The realization example we analyzed is working in the regime of huge error probability, $p \approx 0.5$, with zero logic threshold voltage and in the sub-noise signal amplitude limit; and all these parameters are very unusual.

Even though that at this stage there are more open questions than answers and we were are at the moment unable to show a functioning thermal noise driven logical circuitry, certain questions can already be answered about it. Estimations were given about the information channel capacity and the minimal energy dissipation $E_1$ given in *Energy/bit* unit. The minimal energy requirement, which is the deterministic energy carried by the signal (which is buried by thermal noise) is given as:

$$\eta = \frac{P_s}{C_{dig}\big|_{p \approx 0.5}} = \frac{\pi \ln 2}{2} kT / bit \approx 1.1 \ bit / kT$$

The main advantage of such a hypothetical thermal noise driven computer would be a potentially improved energy efficiency and an obvious lack of leakage current, cross-talk and ground EMI problems due the very low DC voltages. An apparent disadvantage is the large number of extra (redundancy) elements required for error reduction [11-13].

From the long list of open questions, we list some of the most important ones:

1. Do we need error correction at all (except input/output operations) in such a computer or when we want to simulate the way the brain works?

2. Is there any way to realize a non-Turing machine with stochastic elements without excessive hardware/software based redundancy?

3. How should redundancy and error correction be efficiently used to run the thermal noise driven computer as a Turing machine?

4. How much energy would such error correction cost?

5. How much energy is needed to feed the active devices processing and producing the sub-thermal noise signals or the bandwidth control of thermal noise?

6. What is the impact of the internal noise of these devices?

Though, all these questions are relevant for the ultimate energy dissipation of thermal noise driven computers, the lower limit given for that specific arrangement with DC signal buried by thermal noise stays valid because this is the minimal energy needed to generate itself the digital signal.

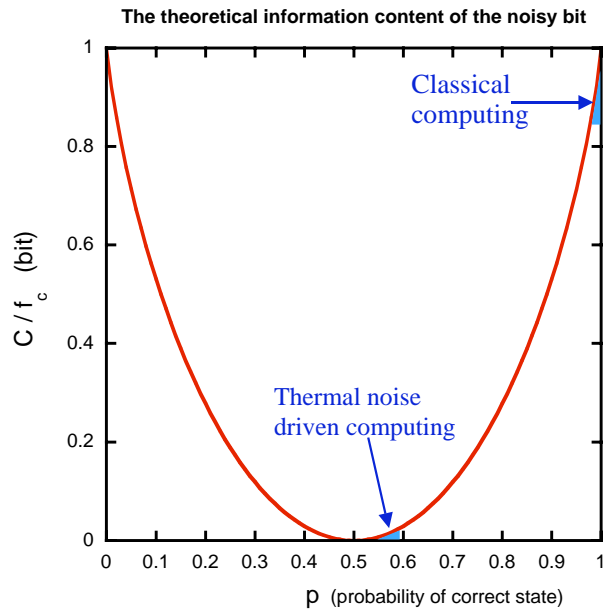

Figure 1. Illustration of the range of error probabilities of classical computing and that of the thermal noise driven computing. A thermal noise driven computer will supposedly work in that range where the information content/ single switching is low, the signal looks like noise and no classical computer could work.

**Introduction.**

Recently, a totally secure classical communication scheme, a statistical-physical competitor of quantum communicators, was introduced, see Figure 1, [1,2], the Kirchhoff-loop-Johnson(-like)-Noise (KLJN) communicator, with two identical pairs of resistors and corresponding Johnson(-like) noise voltage generators. The KLJN communicators are utilizing the physical properties of an idealized Kirchhoff-loop and the statistical physical properties thermal noise (Johnson noise). The resistors (low bit = small resistor, high bit = large resistor) and their thermal-noise-like voltage generators (thermal noise voltage enhanced by a pre-agreed factor) are randomly chosen and connected at each clock period at the two sides of the wire channel. A secure bit exchange takes place when the bit states at the two line ends are different, which is indicated by an intermediate level of the *rms* noise voltage on the line, or that of the *rms* current noise in the wire. The most attractive properties of the KLJN cipher are related to its security [1,3] and to the extraordinary robustness of classical information when compared that to the fragility of quantum information. To provide security against arbitrary types of attacks, the instantaneous currents and voltages are measured at both end by Alice and Bob and they are published and compared. In the idealized scheme of the KLJN cipher, the passively observing eavesdropper can extract zero bit (zero-bit security) of information and the actively eavesdropping observer can extract at most one bit before getting discovered (one-bit security) [1]. The system has a natural zero-bit security against the *man-in-the-middle attack* which is a unique property among secure communicators [3]. The KLJN system has recently became network-ready [4]. This new property [4] opens a large scale of practical applications because the KLJN cipher can be installed as a computer card [4], similarly to Eternet network cards. Other practical advantages compared quantum communicators are the high speed, resistance against dust, vibrations, temperature gradients, and the low price [1]. It has recently been shown [5] that the KJLN communicator may use currently used wire lines, such as power lines, phone lines, internet wire lines by utilizing proper filtering methods.

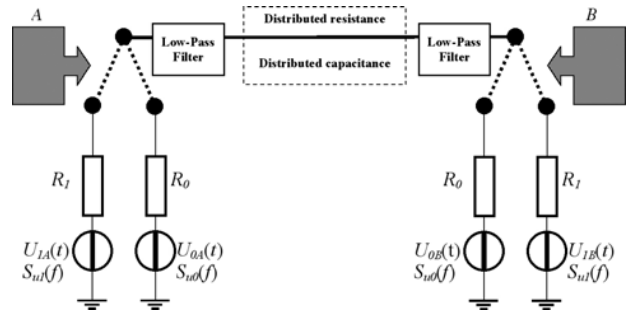

Figure 1. The outline of the KLJN communicator. The line below 2000 km in the relevant frequency range can be represented by distributed resistors and capacitors and below 200 km as a simple resistor. The instantaneous current and voltage data are measured and compared at the two ends and, in the case of deviance; the eavesdropping alarm goes on and the currently exchanged bit is not used. The low-pass line filters are against necessary to protect against out-of-alarm-frequency-band breaking attempts and false alarms due to parasite transients. False alarms would occur due to any wave effect (transient or propagation effects), illegal frequency components or external disturbance of the current-voltage-balance in the wire.

Concerning the security of the KLJN system, the basic rule of any physical secure communication holds here, too: *If you compromise it you lose it*. Though there have been several attempts to break in the KJLN line, so far, no proposed method has been able to challenge the total security of the idealized KLJN system. All these breaking attempts have used certain assumptions which are directly violating the basic model of the idealized KLJN method. These assumptions are as follows:

*i.* Allowing non-negligible wire resistance (Bergou [2], Scheuer and Yariv [6]). For the evaluation of this claim at practical conditions, see the response in [7].

*ii.* Allowing high-enough bandwidth for wave (transient propagation) effects (Scheuer and Yariv [6]). This possibility has been explicitly excluded since the original article [1]. For further refusal of this claim, see the response in [7]. Moreover, *and this is the most direct practical argument*, in this case the standard current and voltage protection of the

KLJN system would *immediately alarm and shut down the communication* because the existence of transient and wave effects *per-definition* yield different instantaneous current and/or voltage values at the two ends.

*iii.* Assuming inaccuracy of (noise) temperatures (Hao [8]). For the theoretical refusal of this claim, see [9].

*iv.* Utilizing practical inaccuracy of resistors (Kish [9]). For the mathematical evaluation of this claim at practical conditions, see there [9].

Let us here consider an analogy in the field of quantum communication. Though all the practical quantum communicators suffer from the inability of producing *totally single-photon* output and this fact is unavoidably compromising the security of practical quantum communicators, we do not say that the idealized/mathematical quantum communicator schemes are insecure. In discussing the unconditional security of idealized quantum communication schemes, we suppose that single-photon sources do exist, which is a similar though more difficult claim than to suppose the existence of zero wire resistance (c.f. superconductors), and then we conclude that the quantum system is therefore unconditionally secure, at least conceptually. Therefore the practical claim that idealized single-photon source does not exists cannot compromise the claim about the security of an idealized, conceptual quantum communicator. On the other hand, as soon as the security of a *practical communicator system* is discussed, the aspects non-ideal single-photon sources, optical fiber absorption, etc, for quantum communicators and similarly the aspects (*i-iv*) listed above for the KLJN systems cannot be neglected any more and the practical design must keep these effects under control.

In the next two sections, we report the first experimental realization and tests of the KLJN secure communicator system. We test and analyze the most important breaking methods of the ones listed above. The tests indicate an unrivalled *beyond-quantum security level* of the realized system.

**2. The realized secure classical physical communicator**

The generic schematic of practical KLJN secure classical communicators is shown in Figure 2. The logic-low (L) and logic-high (H) resistors, $R_0$ and $R_1$, respectively, are randomly selected beginning of each clock period and are driven by corresponding Johnson-like noise voltages. The actual realizations contain more filters and amplitude control units (not shown here). The thick arrows mark computer control and data exchange. The controlling computer (not shown) has a regular network connection with the computer of the other KLJN communicator (not shown).

The realized pair of KLJN communicators [1-4] was designed and built for variable ranges. The circuit realization used Digital Signal Processor (DSP) and analog technology, see the outline in Figures 2 and 3. The details not described here are part of intellectual property which we cannot disclose at this stage. The computer control parts of the communicator pair have been realized by ADSP-2181 type Digital Signal Processors (DSP) (Analog Devices). The digital and analog units were placed on two separate computer cards for easier variability during tuning up. The communication line current and voltage data were measured by (Analog Devices) AD-7865 type AD converters with 14 bits resolution from which 12 bits were used. The DA converters were (Analog Devices) AD-7836 type with 14 bits resolution. The Johnson-like noise was digitally generated in the Gaussian Noise Generator unit where digital and analog filters truncated the bandwidth in order to satisfy the KLJN preconditions of removing any spurious frequency components. The major bandwidth setting is provided by an 8-th order Butterworth filter with sampling frequency of 50 kHz. The remaining small digital quantization noise components are removed by analog filters.

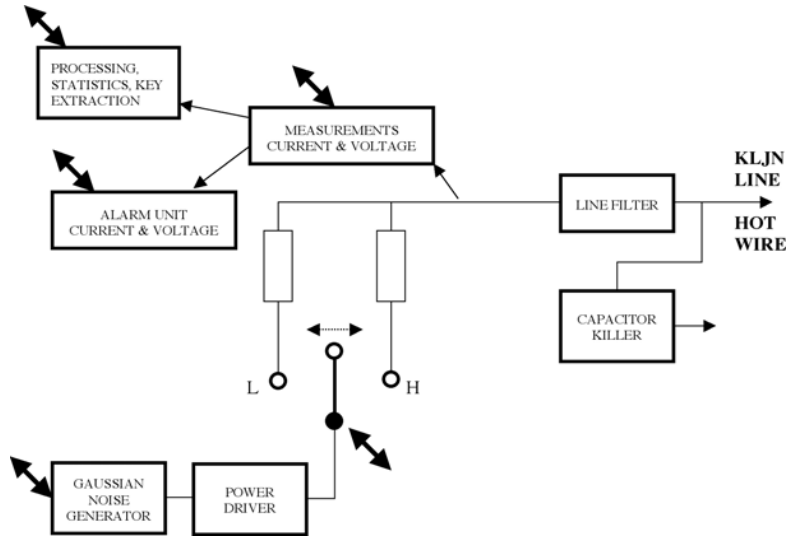

Figure 2. Generic schematics of practical KLJN secure classical communicators. The low (L) and high (H) resistors are randomly selected beginning of each clock period and are driven by corresponding Johnson-like noise voltages. The actual realizations contain more filters and amplitude control units (not shown here). The thick arrows mark computer control and data exchange. The controlling computer (not shown) has a regular network connection with the computer of the other KLJN communicator (not shown). The capacitor killer can be avoided for distances up to about 2000 km provided the cables are not coaxial ones but they are two free hanging wires with a few meters separation, like power lines.

The experiments were carried out on a model-line, with assumed cable velocity of light of $2*10^8$ m/s, with ranges up to 2000 km, which is far beyond the range of direct quantum channels, or of any other direct communication method via optical fibers. The device has bit rates of 0.1, 1, 10, and 100 bit/second for ranges 2000, 200, 20 and 2 km, respectively.

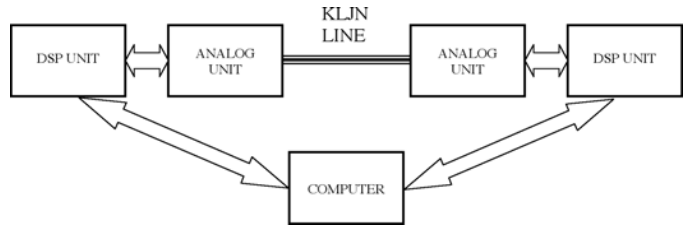

Figure 3. The schematics and arrangement of the realized and tested KLJN communicator pair. The KLJN line is a model line with capacitance compensation up to 2000 km range.

The voltage noises are band-limited Johnson-like noise with power density spectrum

$$S_u(f) = KR \qquad (1)$$

within the noise bandwidth and zero elsewhere. Wave effects in the stationary mode and information leak due to them are avoided by the proper selection of the noise bandwidth. The noise bandwidth is selected so that the highest possible Fourier component in the line is at frequency 10 times lower than the lowest frequency standing-wave mode in the line. That condition results in noise bandwidths 5, 50, 500 and 5000 Hz for ranges 2000, 200, 20 and 2 km, respectively. This condition and the statistical sample size within the clock period determine the speed (bandwidth) of communication. The sampling rate satisfies the Shannon limit thus the sampling frequency is two times greater than the noise bandwidth. The sample size within the clock period is equivalent to about statistically independent 50 data points (though the digital filtering is using about an order of magnitude greater actual sampling rate) and that results in a 50 times lower secure bit rate than the noise bandwidth because only 50% of the communicated bits is secure. The resistors pairs at Alice and Bob are $R_0 = 2$ kOhm and $R_1 = 11$ kOhm.

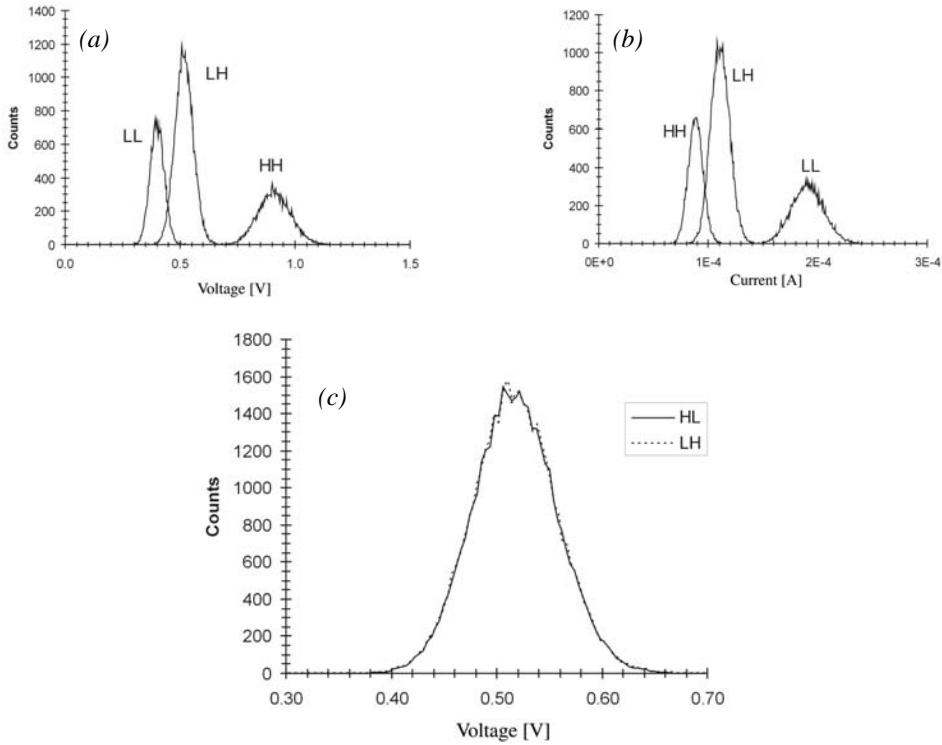

Figure 4. Empirical amplitude density functions (histograms) of the voltage (left window) and current (right window) during the whole span of security checks utilizing 74497 clock cycles: (a) and (b) Voltage and current counts seen by Alice and Bob; (c) Voltage counts seen by Eve one end of the line at the two different secure bit arrangements (*LH* and *HL*). These functions correspond to the situation when the bit arrangement is fixed (*LH* or *HL*) and then the two distribution functions are the voltage counts measured at the two ends of the line to execute a BSchY type of attack. It is obvious from the strong overlap of the two curves that Eve has virtually zero information even with fixed bit arrangement for 74497 clock cycles.

Because the security protection based on current and voltage comparison was effective up to 50 kHz bandwidth, 1 nF capacitors at the two ends of the line were satisfactory *line filters*. Furthermore, these capacitors would have removed possible switching spikes originating from capacitive coupling in the analog switches due to possibly unbalanced parasitic capacitors; therefore there were no detectable switching transients in the line. The 11 kOhm resistor is composed by connecting a 9 kOhm serial resistor to the 2 kOhm resistor. The 2 kOhm resistors are two serial 1kOhm resistors with a 1 nF capacitor shunting their joint point to the ground to remove possible digital quantization noise. The 1 kOhm resistor at the generator dive end was also used as a probe to measure the current in the line. The value of *K* is selected so that the noise voltage of the greater resistor is 1 Volt for all noise bandwidths. This resulted in $S_u(f)$ values of the greater resistor 0.2, 0.02, 0.002, 0.0002 $V^2/Hz$ for ranges 2000, 200, 20 and 2 km, respectively.

Transient wave effects at the end of clock period are avoided in the Gaussian Noise Generator unit by driving the envelope of the time functions of noise voltage and current to zero before the switching using a linear ramp amplitude modulation (via 8% of the clock duration); and the reverse process is done at the beginning of the next clock cycle after the switching of resistors. Moreover a short pause (8 % of the clock time) with no data collection, except for security check, after the initial linear ramp at the beginning of stationary noise, is applied in order to avoid possible other types of transient effects of stochastic nature (though we have not seen any transients). All these are done before the filtering process to avoid any spurious frequency components due to the linear ramp.

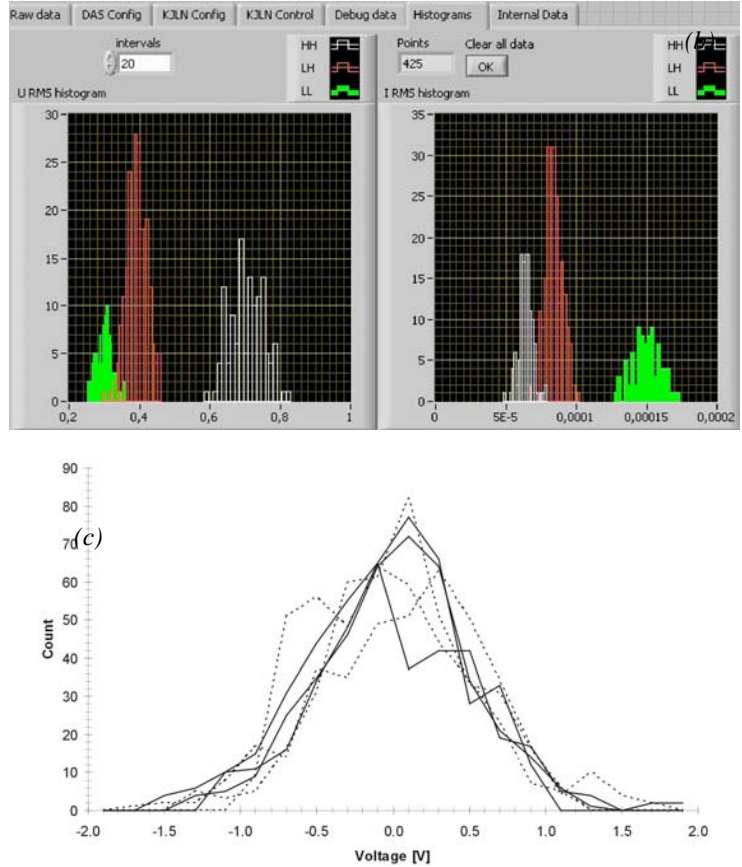

Figure 5. Empirical amplitude density functions (histograms) of the voltage (left window) and current (right window) during a single clock cycle of the communication: (a) and (b) Voltage and current counts seen by Alice and Bob; (c) Voltage counts seen by Eve at one end of the line at the two different secure bit arrangements (dashed curves: 3 independent records of *LH*; solid curves: 3 independent records of *HL*). They correspond to about 50 independent samples though the number of sampled points is greater. These histograms in figure (c) correspond to the situation when the bit arrangement is fixed (LH or HL) and then the two distribution functions are the voltage counts measured at the two ends of the line to execute a BSchY type of attack. The poor statistics seen in figures (a) and (b) are enough for Alice and Bob to identify secure bit alignment with 0.02% error rate (99.88% fidelity). However when Eve tries to identify the bits from the two histogram recorded at the two ends of the line (see figure (c)) she must work with these distributions which are very stochastic, almost identical and totally overlapping with a 1% or less shift of their centers [7] which results in less than 0.19% eavesdropped bit / transmitted secure bit.

The wire diameters of the model line are selected so that they resulted in about 200 Ohm internal resistance for all the different ranges. The corresponding copper wire diameters are reasonable practical values for the different ranges are 21 mm (2000 km), 7 mm (200 km), 2.3 mm (20 km) and 0.7 mm (2 km). Inductance effects are negligible with the selected resistance values, $R_0$ and $R_1$, at the given ranges and the corresponding bandwidths. If the wire is a free hanging one with a few meters separation from earth, such as power lines, parasitic capacitances are not a problem either. However, if the wire is a coaxial cable then with the present driving resistances $R_0$ and $R_1$, a capacitor killer application is needed. The capacitor killer is a well-know electronic solution to remove capacitive cut-off effects in coaxial cables by feeding the coat of a coaxial cable by a follower driven by the hot wire in the coax cable. Therefore, for the elimination of the cable capacitance effects, either a free-hanging wire or a coaxial cable with capacitor killer arrangement was supposed.

## 4. 3. Experimental study of the security and the fidelity of the realized communicators

During the communication and security tests, the exchange of secure bits took place between Alice and Bob with 0.02% error rate, which is 99.98% fidelity, much beyond that of quantum channels. Figure 4 shows an illustrative example about the actual distribution functions and their statistics made over the over the total span of security checks (about 74497 clock cycles) and Figure 5 shows an example of the actual counts (working statistics) during a single clock cycle.

According to the security tests and the related evaluations the raw-bit security of this practical system has been successfully designed to outperform raw-bit quantum security while it still remain reasonably inexpensive. For the summary of security test evaluation, see Table 1.

Breaking tests *(i)* of Bergou-Scheuer-Yariv (BSchY) type utilizing wire resistance [2, 6] were carried out by measuring and comparing the *rms* voltages at the two ends of the line and, during secure bit transfer, supposing *LH* or *HL* bit arrangement, accordingly. We have been using Shannon's channel coding theorem is the tool used to evaluate the bit leak of this kind [7]:

$$\frac{C_{eav}}{C_{trans}} = 1 + p\log_2 p + (1-p)\log_2(1-p) \qquad (2)$$

where $C_{eav}$ is the upper limit of eavesdropped bits (supposing optimal coding/decoding), $C_{trans}$ is the total number of transferred secure bits, $p$ is the probability of correct assumption and $(1-p)$ is the probability of erroneous assumption. The experimental test of this practical information leak through 74497 clock cycles shows $p = 0.526$ which yields $C_{eav} / C_{trans} = 0.19\%$ and that is less than practical quantum bit-leak when photons are randomly stolen, copied (noisy cloned) and replaced; a method by which Eve can easily extract 1% of the bits while staying hidden at practical situations. However to reach this 0.19% information leak Eve must know/use the optimal decoding of data, in accordance with Shannon's channel coding theorem, and that is usually unknown. Moreover, the location of correctly eavesdropped bits is random and that is also a significant deficiency as compared to the random stealing of photons in quantum channels where the location of the correctly eavesdropped bit is known (with reasonable accuracy). A simple improvement possibility is served by the fact that the BSchY bit-leak scales with the reciprocal of the sixth power of cable diameter [10] and that means that doubling cable diameter would reduce this 0.19% leak by a factor of 64 and a diameter increase of a factor of 10 (possible only for 20 km or shorter ranges due to practical limits of cable diameter) would yield a million times reduction of the bit leak. Note, using additional software tools can provide arbitrarily strong reduction of the bit-leak; this is the *privacy amplification* technique [12,13] developed for quantum communication by Bennett and coworkers. However, the price of that is slowing down.

| TYPE OF BREAKING | MEASURED NUMBER, OR RATIO, OF EAVESDROPPABLE BITS WITHOUT SETTING ON THE CURRENT-VOLTAGE ALARM (TESTED THROUGH 74497 BITS) | REMARKS |
|---|---|---|
| BSchY (*i*) [2,6] attack in the present KLJN system | 0.19% | 0.00000019% at 10 times thicker wire (theoretical extrapolation). Arbitrarily can be enhanced by privacy amplification [12,13]; the price is slowing down. |
| Hao (*iii*) [8] attack in the present KLJN system | Zero bit | Below the statistical inaccuracy. Considering the 12 bit effective resolution of noise generation accuracy, it is theoretically: |
| | | < 0.000000006% |
| Kish (*iv*) [9] attack utilizing resistor inaccuracies in the present KLJN system | Zero bit | Below statistical inaccuracy. Theoretically, when pessimistically supposing 1% resistance inaccuracy, it is: < 0.01% |
| Current pulse injection (Kish) [1] in the present KLJN system | Zero bit | One bit can be extracted while the alarm goes on thus the bit cannot be used. |
| Comparison with breaking into a quantum channel by capturing and multiplying (noisy cloning) each photons and restoring them in the line. | 1000 - 10000 bits | Because only the change in the detection error rate is able to uncover Eve. And that needs to build very good statistics. |
| Comparison with breaking into a quantum channel by randomly stealing photons, multiplying (noisy cloning) and feeding back to the line. | >1% | Arbitrarily can be enhanced by privacy amplification [12,13]; the price is slowing down. |

**Table 1.** Summary of the security analysis of the realized KLJN communicator pair and a quick comparison with quantum security.

Hao's *(iii)* breaking method yields zero (non-measurable) information for the eavesdropper due to the 12 bits effective accuracy of current and voltage measurement and the limited statistics during the clock period. Considering the 12 bit effective resolution of noise generation accuracy, theoretical estimation based on Eq. (2), see [7,9], yields an information leak of less than 0.000000006%.

Kish *(iv)* [9] attack utilizing resistor inaccuracies in the present KLJN system resulted zero extractable information because of the accurate choice of resistors at the two line ends. Supposing 1% resistance inaccuracy, the bit leak is less then 0.01%, according to estimations based on Eq. (2), see [7,9].

Current pulse injection breaking attempts (Kish) [1] were performed by discharging a capacitor on the line to inject a short current pulse and use the current distribution to read out the resistance values. These breaking attempts are able to extract just a single bit of information while the current-alarm goes on. Therefore, as expected, this type of break yields zero useful bit extraction. The smallest detectable current difference at the two ends of the line current was 0.025% of the *rms* current in the line. Though a much less sensitivity would have produced the same practical security level, due to the given resolution of the AD converters, we tested with this accuracy.

Finally, we would like to note that, because breaking tests by the *man-in-the-middle attack* require involved hardware and significantly more efforts, they will be performed later.

5. 4. Conclusion

The realized KLJN communicator shows excellent performance through the model-line through communication ranges orders of magnitude beyond the ranges of known direct quantum communication channels. While security claims of quantum communicators are usually theoretical because of the expensive nature of relevant breaking tests, in our practical system we have been able to carry out extensive security tests with various ways of breaking attempts. The study indicates that the most important type of security leak, which will need most of resources to control, is the Bergou-Scheuer-Yariv type wire resistance effect [2,6,7]; and the rest of the practical security compromises can be neglected, such as the transient/wave arguments of Scheuer-Yariv [6,7], the noise strength argument of Hao [8] and the resistor inaccuracy argument of Kish [9]. The results indicate unrivalled fidelity and security levels among existing physical secure communicators and there are straightforward ways to further improve security, fidelity and range, if it is necessary and resources are available, such as thicker cable, capacitor killer arrangement, etc.

**Acknowledgements**

Zoltan Gingl is grateful for the Bolyai Fellowship of Hungarian Academy of Sciences. The travel of LBK to the University of Szeged for the startup phase of the experiments was covered by the Swedish STINT foundation and the cost of staying (10-15 December, 2006) was partially covered by the European Union's SANES grant. The costs of the KLJN system design were partially covered by the TAMU Information Technology Task Force (TITF, grant 2002). Part of these results is also being submitted to Physics Letters A [14].

**References**

1. [1] L.B. Kish, "Totally secure classical communication utilizing Johnson (-like) noise and Kirchhoff's law", *Physics Letters A* **352** (2006) 178-182.
2. [2] Adrian Cho, "Simple noise may stymie spies without quantum weirdness", *Science* **309** (2005) 2148.
3. [3] L.B. Kish, "Protection against the man-in-the-middle-attack for the Kirchhoff-loop-Johnson(-like)-noise cipher and expansion by voltage-based security", *Fluctuation and Noise Letters* **6** (2006) L57-L63.
4. [4] L.B. Kish and R. Mingesz, "Totally secure classical networks with multipoint telecloning (teleportation) of classical bits through loops with Johnson-like noise", *Fluctuation and Noise Letters* **6** (2006) C9-C21.
5. [5] L.B. Kish, "Methods of Using Existing and Currently Used Wire Lines (power lines, phone lines, internet lines) for Totally Secure Classical Communication Utilizing Kirchoff's Law and Johnson-like Noise", draft, http://arxiv.org/abs/physics/0610014 .
6. [6] J. Scheuer, A. Yariv, "A classical key-distribution system based on Johnson (like) noise—How secure?", *Physics Letters A* **359** (2006) 737.
7. [7] L.B. Kish, "Response to Scheuer-Yariv: "A Classical Key-Distribution System based on Johnson (like) noise - How Secure?"", *Physics Letters A* **359** (2006) 741–744.
8. [8] F. Hao, "Kish's key exchange scheme is insecure", *IEE Proceedings on Information Security* **153** (2006) 141-142.
9. [9] L.B. Kish, "Response to Feng Hao's paper "Kish's key exchange scheme is insecure"", *Fluctuation and Noise Letters* **6** (2006) C37–C41.
10. [10] L.B. Kish, R. Mingesz and Z. Gingl, "How Secure, How to Design, and How to Run Practical Kirchhoff-loop-Johnson-like-Noise Ciphers - I Theory", in preparation.
11. [11] R. Mingesz, Z. Gingl and L.B. Kish, "How Secure, How to Design, and How to Run Practical Kirchhoff-loop-Johnson-like-Noise Ciphers - II Practice", in preparation.
12. [12] C.H. Bennett, G. Brassard, C. Crepeau and U.M. Maurer, "Generalized Privacy Amplification", *IEEE Transactions on Information Theory* **41** (199) 1915-1923.
13. [13] A. Berzanskis, "Applications of Quantum Cryptography in Government Classified, Business, and Financial Communications", Supercomputing'05, Seattle November 12-18, 2005.
14. [14] L.B. Kish, R. Mingesz, Z. Gingl, "Johnson(-like)-Noise-Kirchhoff-Loop Based Secure Classical Communicator Demonstrated for Ranges of Two to Two Thousand Kilometers", Physics Letters A, submitted (April 16th, 2007).